\documentclass[a4paper!]{elsart}

 \usepackage{graphicx}

\usepackage{amssymb}

 \usepackage{lineno}
\usepackage{subfigure}

\newcommand{\zpt}{\mbox{$Q_T$}}
\newcommand{\zptw}{\mbox{$Q_T^*$}}
\newcommand{\dphi}{\mbox{$\Delta\phi^{ll}$}}
\newcommand{\at}{\mbox{$a_T$}}
\newcommand{\al}{\mbox{$a_L$}}
\newcommand{\bt}{\mbox{$b_T$}}
\newcommand{\bl}{\mbox{$b_L$}}
\newcommand{\zptgen}{\mbox{$Q_{T gen}$}}

\newcommand\T{\rule{0pt}{2.6ex}}
\newcommand\B{\rule[-1.2ex]{0pt}{0pt}}

\newcommand{\ptone}{\mbox{${\vec{p}_T}$$^{(1)}$}}
\newcommand{\pttwo}{\mbox{${\vec{p}_T}$$^{(2)}$}} 

\newcommand{\zmumu}{\mbox{$Z \rightarrow \mu^+\mu^-$}}
\newcommand{\vecqt}{\mbox{$\vec{Q}_T$}}
\begin{document}

\begin{frontmatter}



\title{A Novel Technique for Studying the \boldmath{$Z$} Boson Transverse Momentum Distribution at Hadron Colliders}


 \author[]{M. Vesterinen, T. R. Wyatt.}

\author{}

\address{Particle Physics Group, School of Physics and Astronomy, University
of Manchester, UK.}

\date{July 30, 2008}
\begin{abstract}
  We present a novel method for studying the shape of the $Z$ boson
transverse momentum distribution, \zpt,
  at hadron colliders in $p\bar{p}/pp \rightarrow Z/\gamma^* \rightarrow l^+l^-$.
  The \zpt\ is decomposed into two orthogonal
components; one transverse and the other parallel to the di-lepton thrust axis.
  We show that the transverse component is
  almost insensitive to the momentum resolution of the
  individual leptons
  and is thus more precisely determined on an event-by-event basis than the
  \zpt.
  Furthermore, we demonstrate that a measurement of the 
  distribution of this transverse component is substantially less sensitive to the dominant
  experimental systematics (resolution unfolding and \zpt\ dependence of
event selection efficiencies)
  reported in previous measurements of the \zpt\ distribution.

\end{abstract}

\begin{keyword}

\PACS 12.38.Qk \sep 13.85.Qk \sep 14.70.Hp \sep 12.38.Lg \sep 12.15.Mm 
\end{keyword}
\end{frontmatter}


\def\plotdir{figures}

  \section{Introduction}    
  The shape of the $Z$ boson momentum distribution transverse to the beam direction (\zpt) at a hadron collider tests the predictions of 
  quantum chromo dynamics (QCD),
  since non-zero \zpt\ is generated through radiation from the initial state partons.
  A good understanding of electroweak vector boson production is
  important in precision measurements (e.g. top quark and $W$ boson mass) and in Higgs boson and new phenomena searches at hadron collider experiments.
  In many of these searches the signal and/or background processes involve electroweak vector bosons produced in association with jets.

  At low \zpt\,~(\zpt\ $\ll Q$), where the emission of multiple soft gluons is important,
  calculations in fixed order perturbative QCD diverge.
  There exist resummation techniques,
  in which singular contributions from all orders of $\alpha_s$ are resummed to give a finite result.
  Resummation was first applied to the Drell-Yan process by Collins, Soper and Sterman (CSS)~\cite{CSS1985}.
  The resummation is carried out in impact parameter ($b$) space and
includes a non-perturbative (NP) form factor that needs to be determined from data.
  Brock, Landry, Nadolsky and Yuan (BLNY) proposed the following parameterisation~\cite{BLNY2003}:

  \begin{equation}
  S_{NP}(b,Q^2) = \left[g_1 + g_2\ln\left(\frac{Q}{2Q_0}\right) +
g_1g_3\ln(100x_ix_j)  \right]b^2
  \label{BLNYeq}
  \end{equation}

where $x_i$ and $x_j$ are the fractions of the hadron momenta carried by
the initial state partons, $Q_0$ is an arbitrary scale
set to $1.6$ GeV and the parameters $g_i$ need to be fitted from data.
 
 Using the BLNY NP form factor, the CSS formalism was able to 
 describe simultaneously Tevatron Run~I $Z$ data and lower $Q^2$ Drell-Yan data in a global
 fit of the parameters $g_i$~\cite{BLNY2003}.
 The \zpt\ distribution at the Tevatron ($Q^2 \sim M_Z^2$) is sensitive to $g_2$ and insensitive to $g_1$ and $g_3$.
 A larger value of $g_2$ corresponds to a harder \zpt\ distribution.
 The CSS formalism is implemented in the next to leading order (NLO) event generator ResBos~\cite{ResBos}.
 In Run~II the D\O\ Collaboration reported a \zpt\ measurement in the di-electron channel 
 with 1~fb$^{-1}$ of data~\cite{DzeroRunIIa}.
 For low \zpt\ ($\zpt < 30$ GeV), the D\O\ data is, within the measurement uncertainties,
 well described by the CSS/BLNY formalism.
   
  At low \zpt\, the overall uncertainties were dominated by the parton distribution functions (PDFs) and the
   following experimental systematics~\cite{DzeroRunIIa}:
  \begin{itemize}
  \item Unfolding the \zpt\ measurement to account for the resolution in the measurement of the $E_T$ of the electrons.
  \item Correcting for the \zpt\ dependence of the overall event selection efficiency.
  \end{itemize}
  As a result of these substantial experimental systematics, the low \zpt\ region was not much better measured in this 1~fb$^{-1}$
  Run~II analysis than in the 100~pb$^{-1}$
  Run~I analysis \cite{DzeroRunI}.
  In both analyses, a measurement was made of $g_2$.
  The measurements: $0.59\pm0.06$ GeV$^2$ for Run~I and $0.77\pm0.06$
  GeV$^2$ for Run~II, have comparable uncertainties~\footnote{
  The D\O\ Run~I and Run~II measurements cannot be directly compared since they
  used different PDFs and the Run~I measurement used the  
  the Ladinsky-Yuan (LY) parameterisation~\cite{LY} of the NP form factor as opposed to the BLNY parameterisation.
  Only the third term in $g_1g_3$ is different and any shift in the fitted $g_2$ is far smaller than the uncertainties on 
  these measurements.}.
  
  An observable that is sensitive to the \zpt, but less sensitive to these experimental systematics would be beneficial.
  In this work, Monte Carlo studies of an optimal observable, $a_T$, are presented.
  The $a_T$ observable has previously been used in the selection of $l^-l^+\nu\bar{\nu}$ final states at LEP by the OPAL collaboration \cite{OPAL-llvv}.
  The UA2 Collaboration used a similar observable, $p_{\eta}^Z$, which we
  refer to as \bt, in a \zpt\ measurement~\cite{UA2}.

  \section{Constructing the Observable}
  
  The measured \zpt\ is highly sensitive to the lepton $p_{T}$ resolution.
  Our goal is to build an observable that is less sensitive to this resolution,
  whilst still sensitive to the \zpt.
  We keep in mind the fact that collider detectors generally have far better angular resolution than calorimeter $E_{T}$ or track $p_{T}$ resolution.

  \begin{figure}[htbp]
    \centering
    \includegraphics[width=0.9\linewidth]{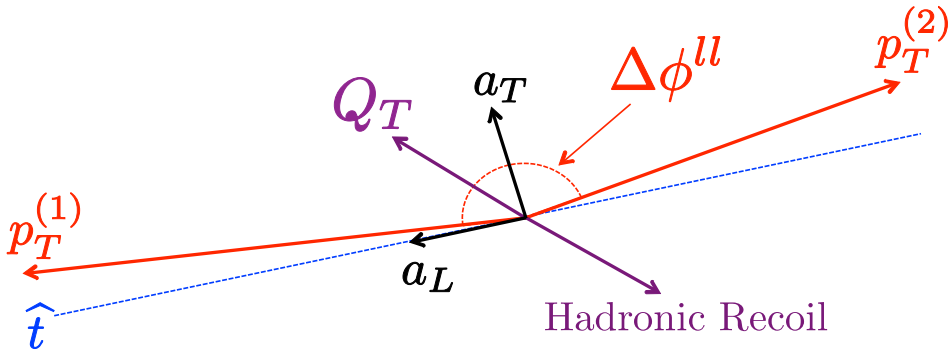}
    \caption{A schematic representation in the transverse plane, of the construction of $a_T$ and $a_L$ in a typical leptonic $Z$ decay.
    The hadronic recoil is expected to have equal and opposite transverse
    momentum to the $Z$.}
    \label{OurObservable}
  \end{figure}
  For events with di-lepton azimuthal separation, \dphi\ $>$ $\frac{\pi}{2}$,
  the \zpt\ is decomposed into orthogonal components as follows (See figure ~\ref{OurObservable}):
  \begin{itemize}
  \item The thrust axis is defined as: 
  $\widehat{t} = \frac{\ptone - \pttwo }{|\ptone - \pttwo|}$
  where $\vec{p}_{T}$$^{(i)}$ is the transverse momentum vector of lepton $i$. 
  The two leptons have equal momentum transverse to this axis.
  \item The transverse momentum vector of the di-lepton system, $\vecqt =
\ptone +\pttwo$, is decomposed into components
    transverse to the axis, $a_T = |\vecqt \times \widehat{t}|$,
    and aligned with the axis, $a_L = \vecqt \cdot \widehat{t}$.
  \end{itemize}
      
  For events with $\dphi < \frac{\pi}{2}$, $a_T$ is set equal to the \zpt,
  while \al\ maintains the same definition for all values of \dphi.
    
  At low \zpt, $\dphi\sim\pi$,
  hence the uncertainty on \at\ is approximately the uncertainty on the individual lepton $p_T$'s 
  multiplied by the sine of a small angle.
  In contrast, the uncertainty on \al\ (and thus also \zpt) is approximately the uncertainty on the individual lepton $p_T$'s
  multiplied by the cosine of a small angle.
  
  An alternative observable is discussed, \bt, whose construction is identical to that of \at\ except that the
  decomposition is performed relative to the di-lepton perpendicular bisector axis: 
  $\widehat{b} = \frac{\ptone - r\pttwo}{|\ptone - r\pttwo|}$
  where $r = |\ptone|/|\pttwo|$.
  The two leptons have equal acoplanarity with respect to this axis.
  In an event in which the leptons have equal $p_T$, the values of \at\ and \bt\ are equal.
  No study is presented of the component longitudinal to the axis, \bl.
  
  As discussed below, the relative sensitivity to lepton $p_T$ mis-measurement of \at\ and
  \bt\ depends on the correlation between the lepton $p_T$ and  
  its resolution.

  \section{Monte Carlo Simulation}
  
  Monte Carlo (MC) events are generated using {\sc pythia}~\cite{PYTHIA},
  which treats at leading order (LO) the process 
  $p\bar{p}\rightarrow Z/\gamma^* \rightarrow \mu^+\mu^-$ plus up to one jet,
   at a centre of mass energy of 1.96 TeV, and mass between 60 and 130~GeV.
   {\sc pythia} uses additional parton showering to simulate ``soft'' transverse momentum generation.
  Although this study is carried out with \zmumu\ events, the idea is
  applicable to studying the di-electron channel.
  
  In order to simulate the imperfect muon  $p_T$ resolution
  of a detector, Gaussian smearing of width 0.003 GeV$^{-1}$ is applied in $1/p_T$ to both muons,
  which is approximately the design muon $p_T$ resolution of the Run~II D\O\
  detector~\cite{DZERO-TDR}.  
  The design tracking resolution of other current and future hadron collider detectors;
  CDF, ATLAS and CMS~\cite{CDF-TDR,ATLAS-TDR,CMS-TDR} vary between $\approx
  10^{-3}$ and $\approx 10^{-4}$ GeV$^{-1}$ 
  depending on the track $p_T$ and pseudo-rapidity ($\eta$). 
  We refer to the constant $\delta(1/p_T)$ form of the resolution $p_T$ dependence as {\it muon-like}.
  An alternative resolution dependence is studied; {\it electron-like}
  resolution with the form:  
  \mbox{$\delta p_T/p_T = 0.15/p^{1/2}$}, which would be the form expected
  for a calorimeter-based measurement appropriate for electrons. 
  Gaussian smearing is applied to the lepton azimuthal angle $\phi$ of width
  0.0005 rad., which is the typical resolution of a detector.

  In order to study the dependence of event selection efficiency on \zpt, \at, \al\ and \bt,  the
  following simple cuts are applied to the event sample.
  These are representative of the cuts typically applied to select $Z$
  decays at a hadron collider.
  \begin{itemize}
  \item The di-muon invariant mass must be between 70 and 110 GeV.
  \item Kinematic cuts on both muons: $p_T > 15$~GeV and $|\eta| < 2$.
  \item An isolation cut on both muons: $f_{iso} < 2.5$ GeV,
    where \mbox{$f_{iso} = \sum p_T(\Delta R < 0.2)$}
    and \mbox{$\Delta R = \sqrt{(\Delta\eta)^2 + (\Delta\phi)^2}$} is the radius of a cone around the
    candidate muon. The sum is over all particles excluding both the candidate muon and 
    any neutrinos.
  \end{itemize}   
  {\sc pythia} is a LO event generator and is not expected to give a good
description of the \zpt\ distribution in real data.
  The events are re-weighted in \zpt\ and $Z$ rapidity ($y$) to match the prediction of ResBos using the
  default settings, which is in good agreement with D\O\ Run~II
data~\cite{DzeroRunIIa} in the region of low \zpt\ (\zpt\ $<$ 40~GeV).
Event samples are generated using ResBos, and the re-weighting procedure is carried out as follows:          

\begin{itemize}   
\item Two
dimensional histograms in \zpt\ and $|y|$ are produced for both the {\sc      
pythia} and ResBos event samples.          
\item Dividing the ResBos histogram by the {\sc pythia}
histogram                                                              
   gives an {\it event-weight} histogram.  
   \item The {\sc pythia} events are now given an {\it event-weight}
   based on                                                             
   their \zpt\ and $|y|$ such that the distributions of these variables are
   the                                                      
   same as the ResBos prediction.
\end{itemize}  

Hereafter, unless otherwise stated, MC refers to {\sc pythia} re-weighted to ResBos.

  \section{Sensitivity to Lepton \boldmath{$p_{T}$} Mis-measurement}
  
  \newcommand{\xgen}{\mbox{$X_{gen}$}}
  \newcommand{\xdet}{\mbox{$X_{det}$}}

  \begin{figure}[htbp]
  \begin{center}
        \subfigure[{\it muon-like}]{\includegraphics[width=0.49\linewidth]{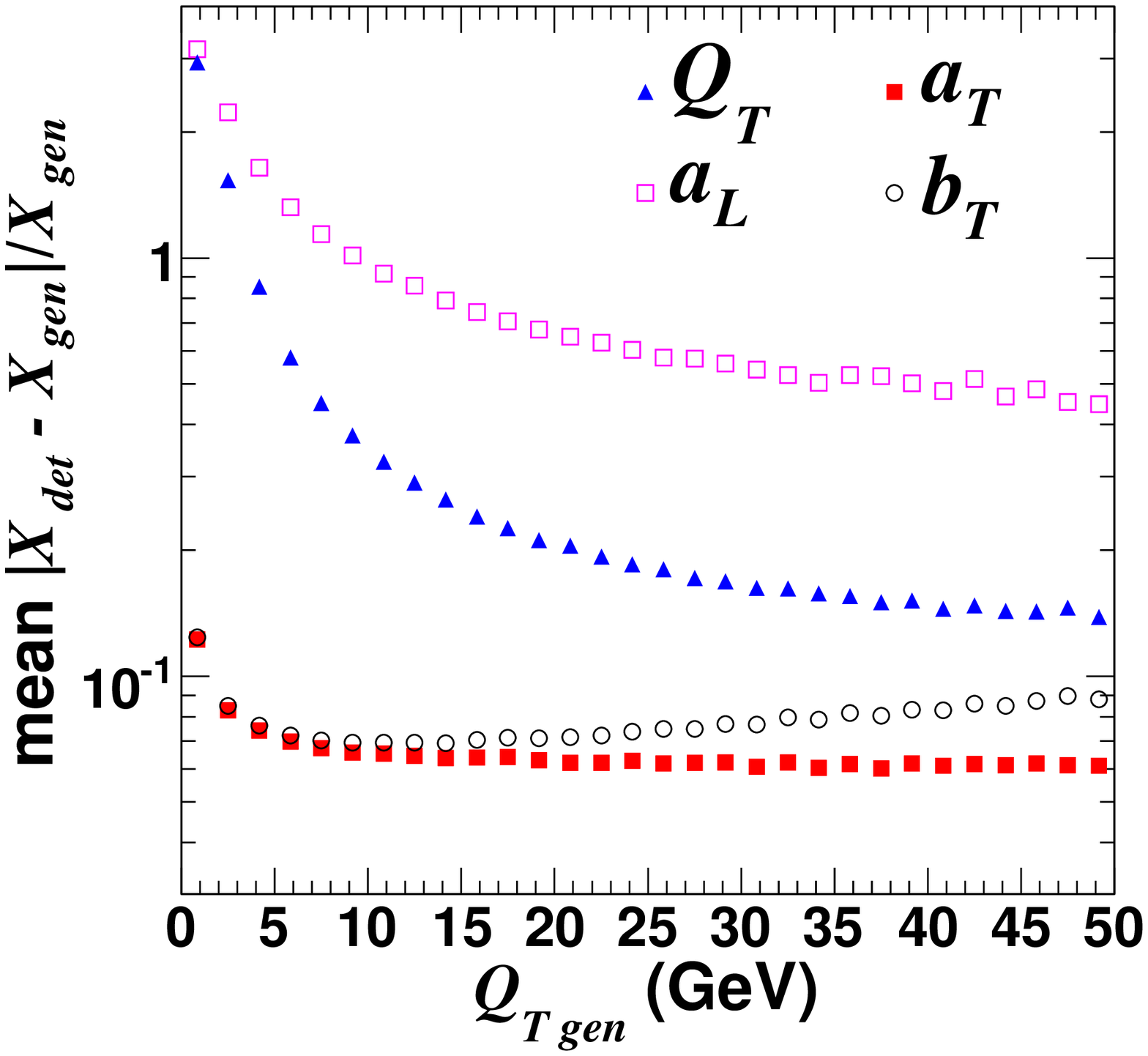}}
        \subfigure[{\it electron-like}]{\includegraphics[width=0.49\linewidth]{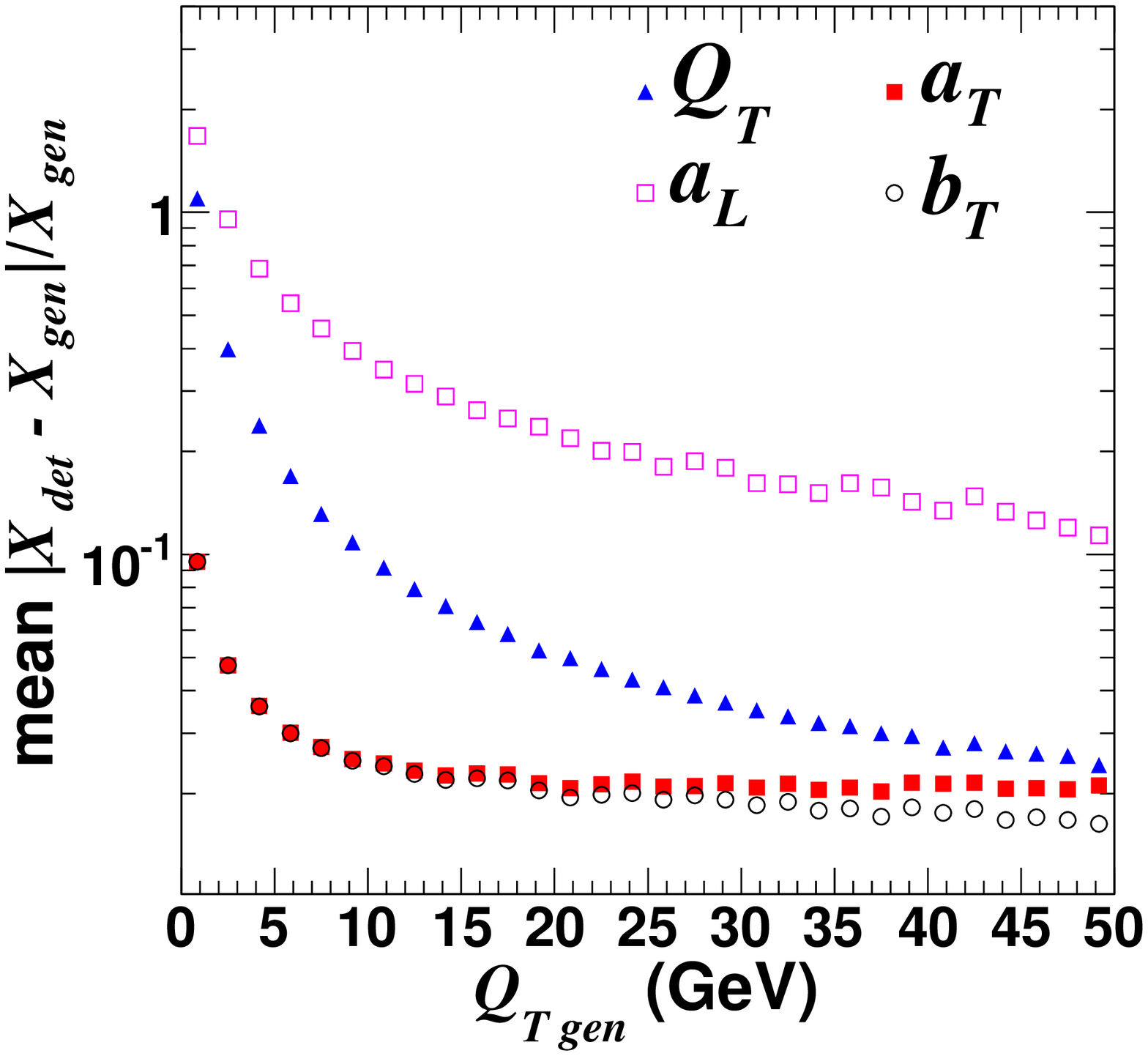}}
           \caption[Caption for the \plotdir/ page]{
           Mean resolution, $|\xdet-\xgen|/\xgen$ for each of the observables, \zpt, \at, \al\ and \bt\ as a function of \zptgen,
	   with (a) {\it muon-like} resolution (\mbox{$\delta (1/p_T) =0.003$ GeV$^{-1}$}) and (b) {\it electron-like} resolution
	   (\mbox{$\delta p_T/p_T = 0.15/p^{1/2}$}).
           }
         \label{resolution}
     \end{center}
 \end{figure}
   
   Two quantities are defined: the smeared or {\it detector level} quantity, \xdet, and 
   the unsmeared or {\it generator level} quantity, \xgen, where $X$ corresponds to \zpt, \at, \al\ or \bt.

  Figures~\ref{resolution}~(a) and ~\ref{resolution}~(b) show the mean resolution,
  \mbox{$|\xdet-\xgen|/\xgen$} as a function of \zptgen\ for {\it muon-like} and
  {\it electron-like} resolution respectively.
  It can be seen that for low to moderate values of \zpt\ ($\zptgen < 50$ GeV), 
  \at\ and \bt\ are significantly better measured than \al\        
  or \zpt.
  Either \at\ or \bt\ are therefore particularly well suited to studying $Z$
  production at low to moderate \zpt.
  In the region of low \zpt\ ($\zptgen < 15$ GeV), \at\ and \bt\ have similar 
  resolutions for either the (a) {\it muon-like} or (b) {\it electron-like} cases.
  At increasing \zpt, \at\ is more suited for (a), and \bt\ more suited for
  (b).
  The construction of \at\ ensures that the acoplanarity angle to the thrust
  axis is smaller for the larger $p_T$ lepton which tends to have poorer
  resolution in the {\it muon-like} case.
  In the following sections, for brevity, we discuss the case of {\it muon-like}
  smearing.

\section{Event Selection Efficiency Dependence}

  \begin{figure}[htbp]
  \begin{center}
    \includegraphics[width=0.49\linewidth]{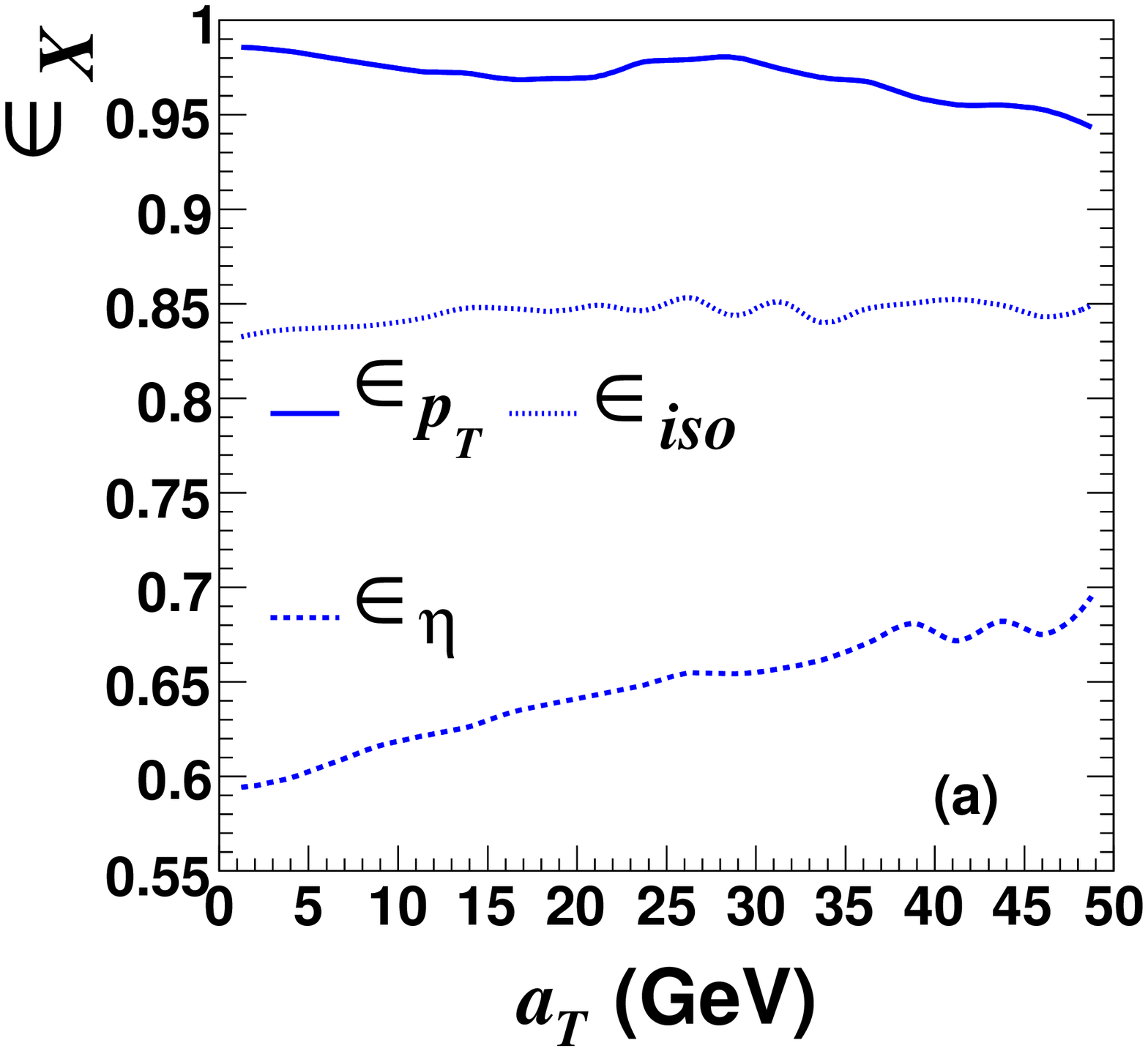}
    \includegraphics[width=0.49\linewidth]{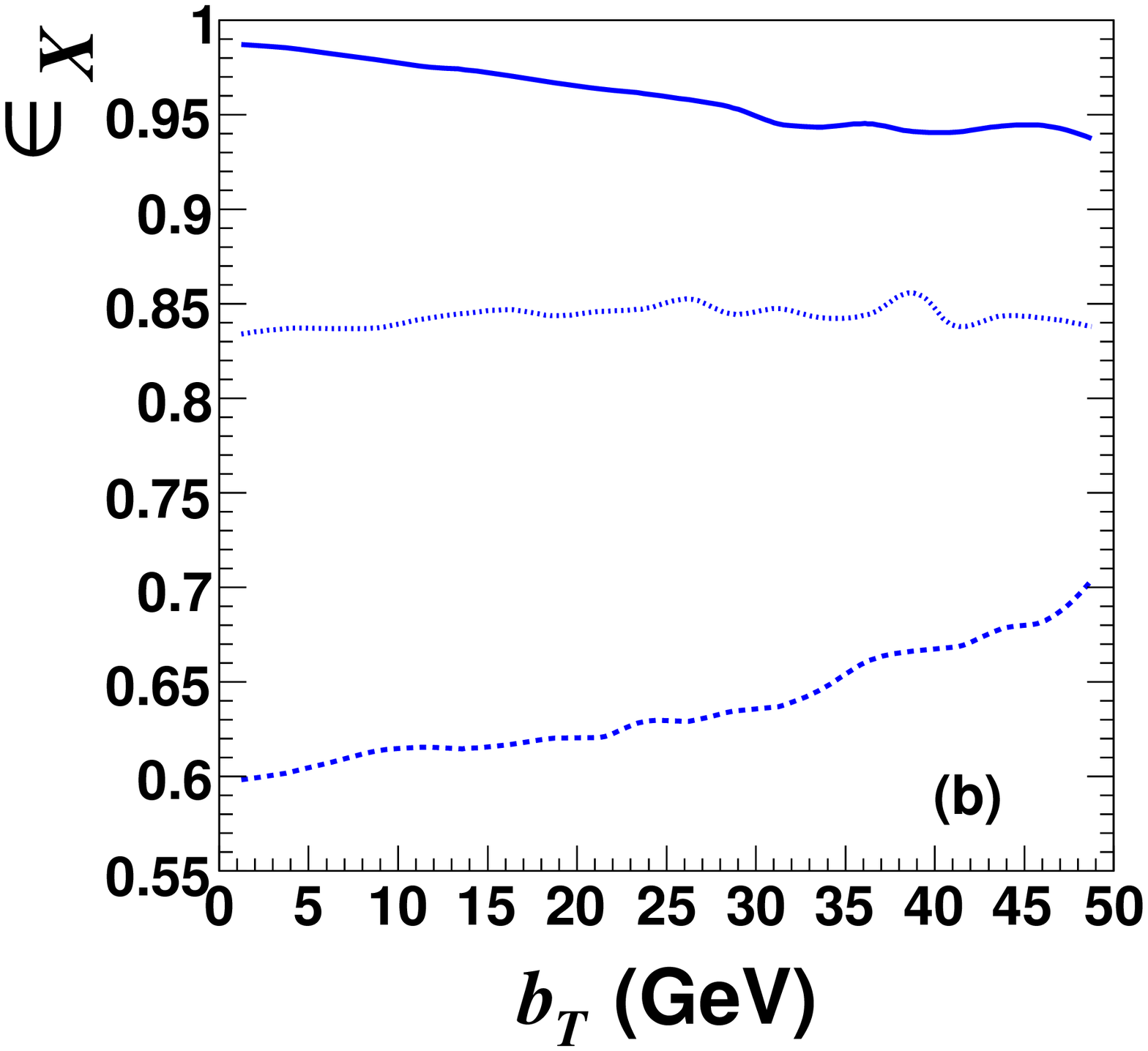}
    \includegraphics[width=0.49\linewidth]{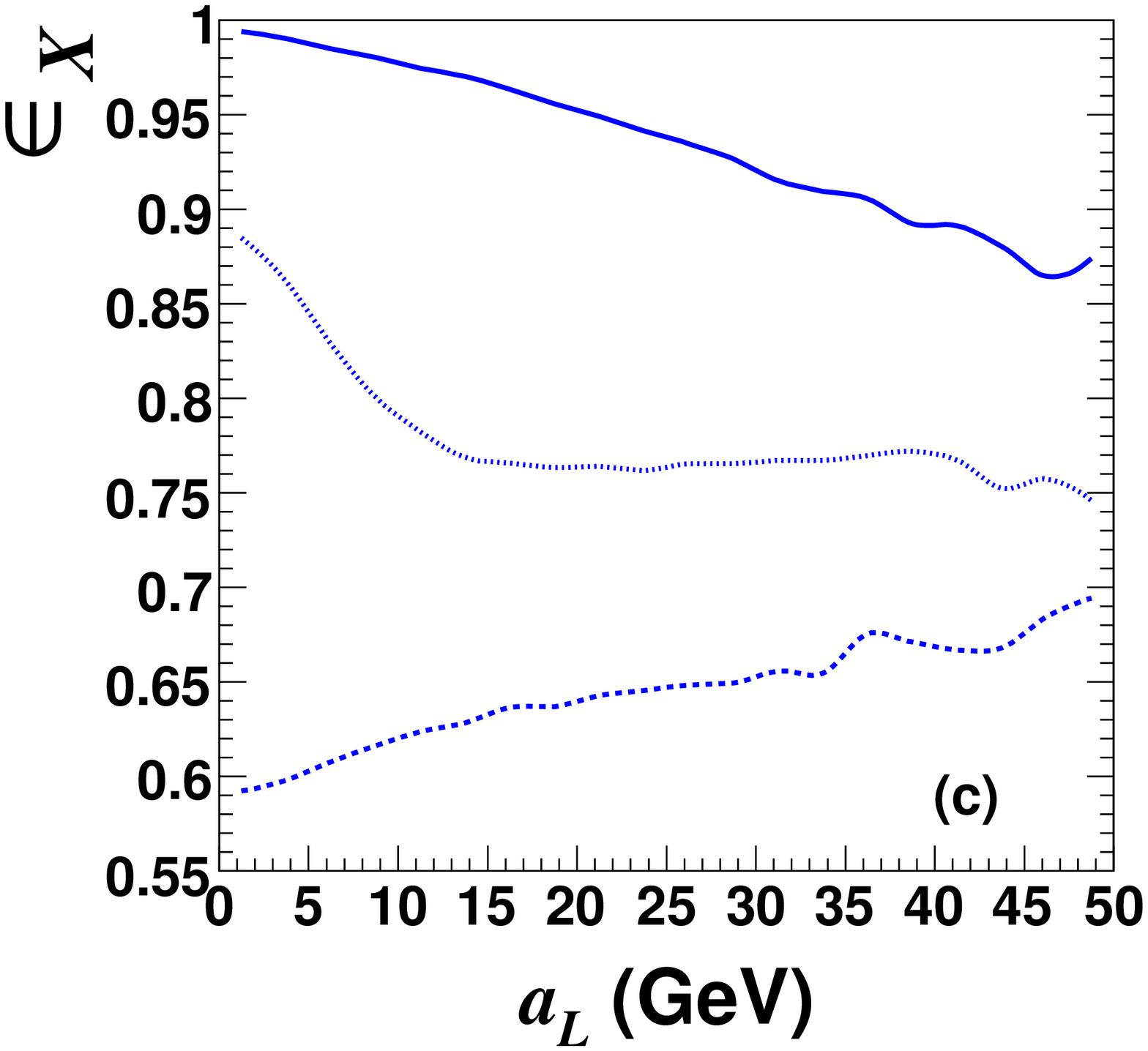}
    \includegraphics[width=0.49\linewidth]{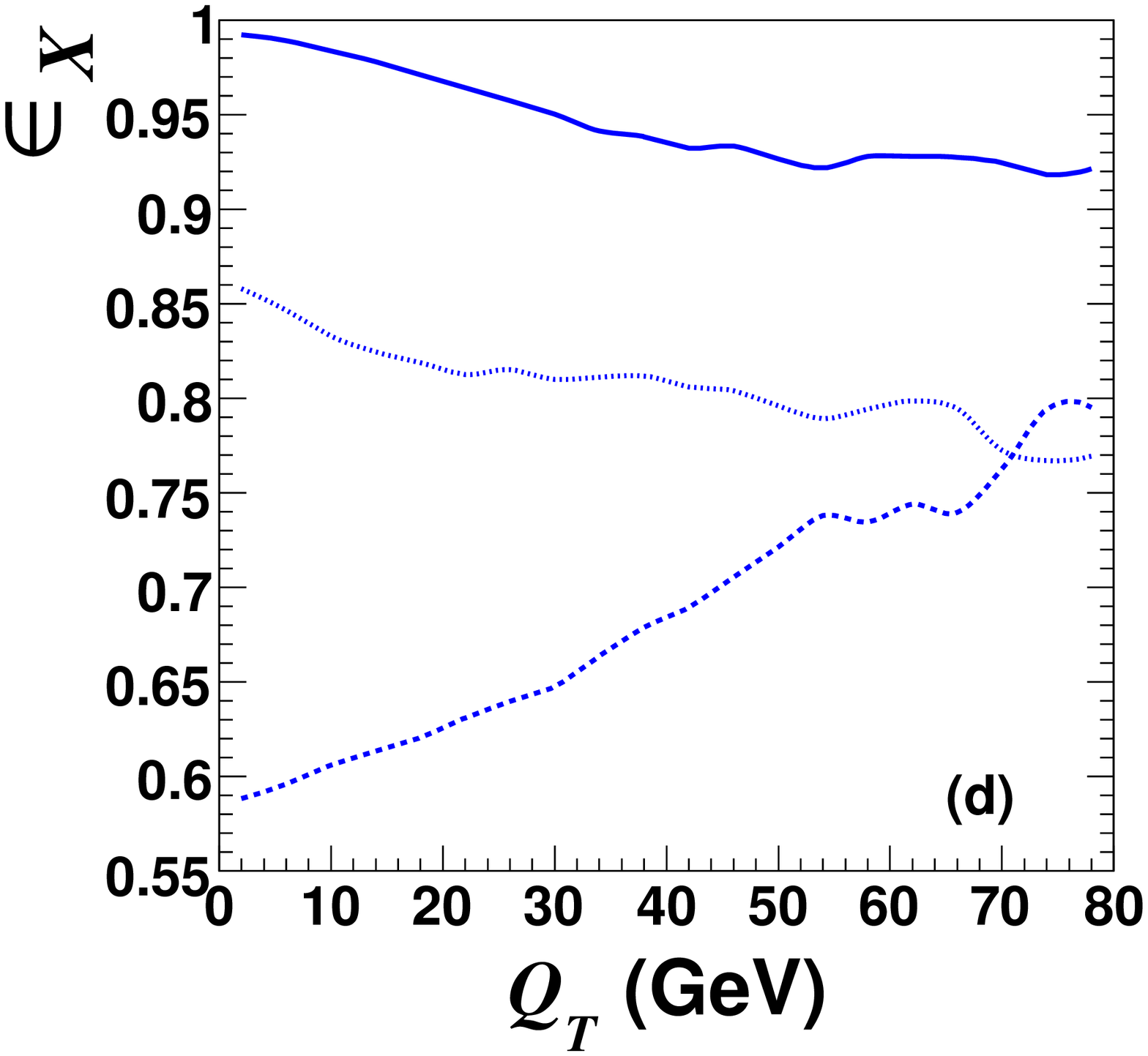}
         \caption{ The dependence of event selection efficiencies on the
generator level (a) \at, (b) \bt, (c) \al\ and (d) \zpt\ in \zmumu\ MC events.
         The efficiencies $\epsilon_{\eta}$, $\epsilon_{p_T}$ and $\epsilon_{iso}$ 
	 (evaluated relative to the previous cut in the order; $|\eta|$, $p_T$, isolation)
	 are shown separately for cuts on
         muon $|\eta|$, $p_T$ and isolation respectively. Note that the
	 range of \zpt\ shown in this figure is $\sqrt{2} \times$
         larger than for $a_T$ and $a_L$. 
         }
         \label{eff_tot_dimu}
	 \end{center}
 \end{figure}

  Figure~\ref{eff_tot_dimu} shows the dependence of the event selection efficiency (separately for the cuts on muon $|\eta|$, $p_T$ and isolation) 
  on the generator level \zpt, \at, \al\ and \bt. 
  The cuts are applied in the following order: $|\eta|$; $p_T$; isolation.
  The efficiency dependence for each cut is calculated having applied all
previous cuts.
  For large \zpt, the muons tend to be more central, increasing the $\eta$ cut efficiency.
  The same correlation is apparent for $a_T$ and $a_L$, although weaker than
  for \zpt.

  The muon $p_T$ cut dependence on $a_T$ is flat in the range considered.
  Conversely large values of $a_L$, generate an asymmetry in the $p_T$'s of the two muons and 
  tend to push the lower $p_T$ muon below the cut threshold.
  The dependence on the isolation cut is substantially flatter for $a_T$ than for $a_L$.
  Large $a_L$ corresponds to a high $p_T$ hadronic recoil with a large fraction of its $p_T$ aligned 
  along the thrust axis and thus possibly lying within the isolation cone of one of the two muons.
  There is no such dependence on $a_T$, since $a_T$ is the component of the recoil $p_T$ transverse to
  the thrust axis.

  In summary, the efficiencies of the cuts on muon $|\eta|$, $p_T$ and
  isolation depend less strongly on \at\ than \zpt.
  The dependence of \bt\ on each of the cuts is similar to that of \at.
    
  \section{Systematic Uncertainties on the Unfolded and Efficiency Corrected Distributions}

  \begin{figure*}[htbp!]
    \centering
    \includegraphics[width=0.49\textwidth]{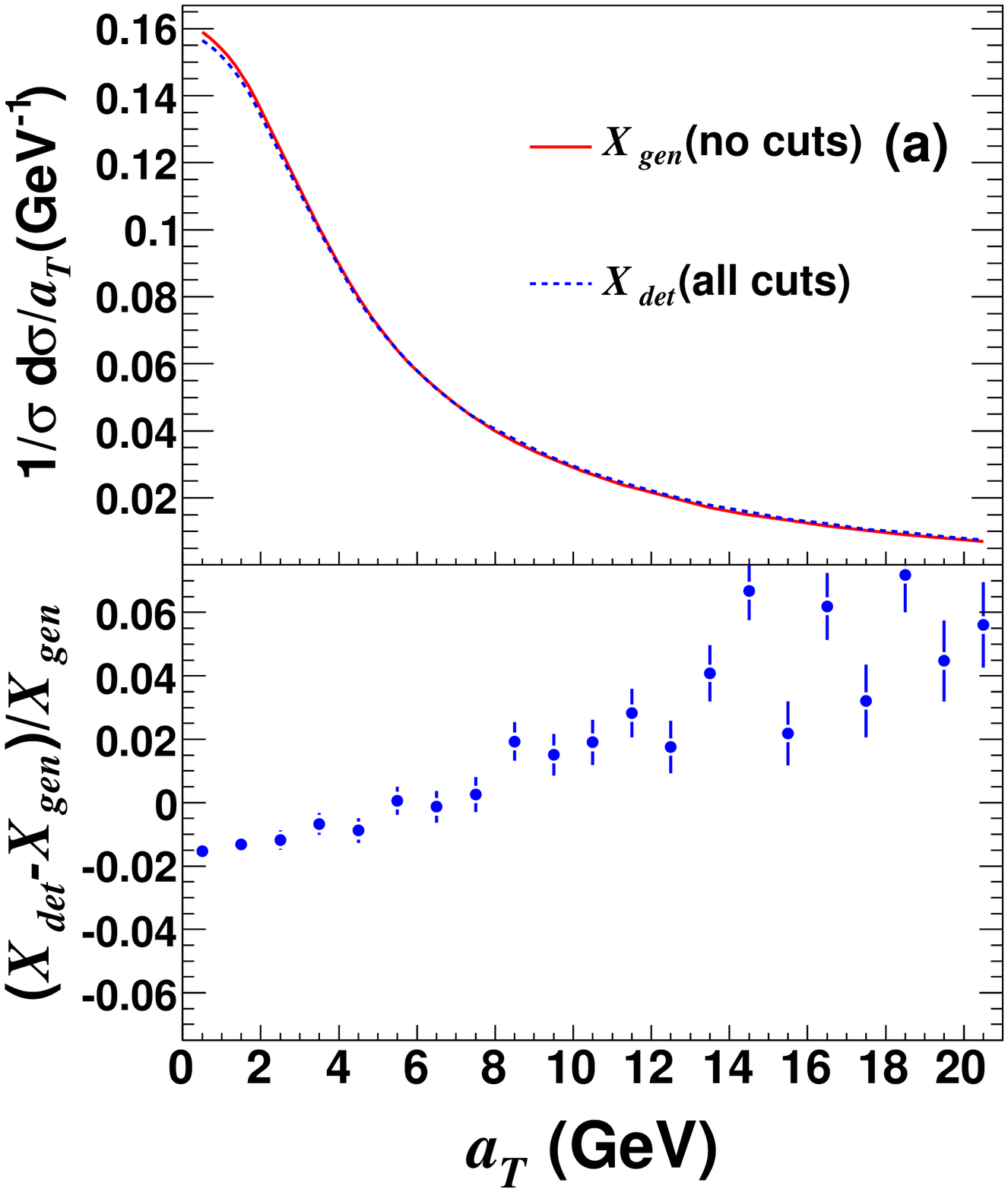}
    \includegraphics[width=0.49\textwidth]{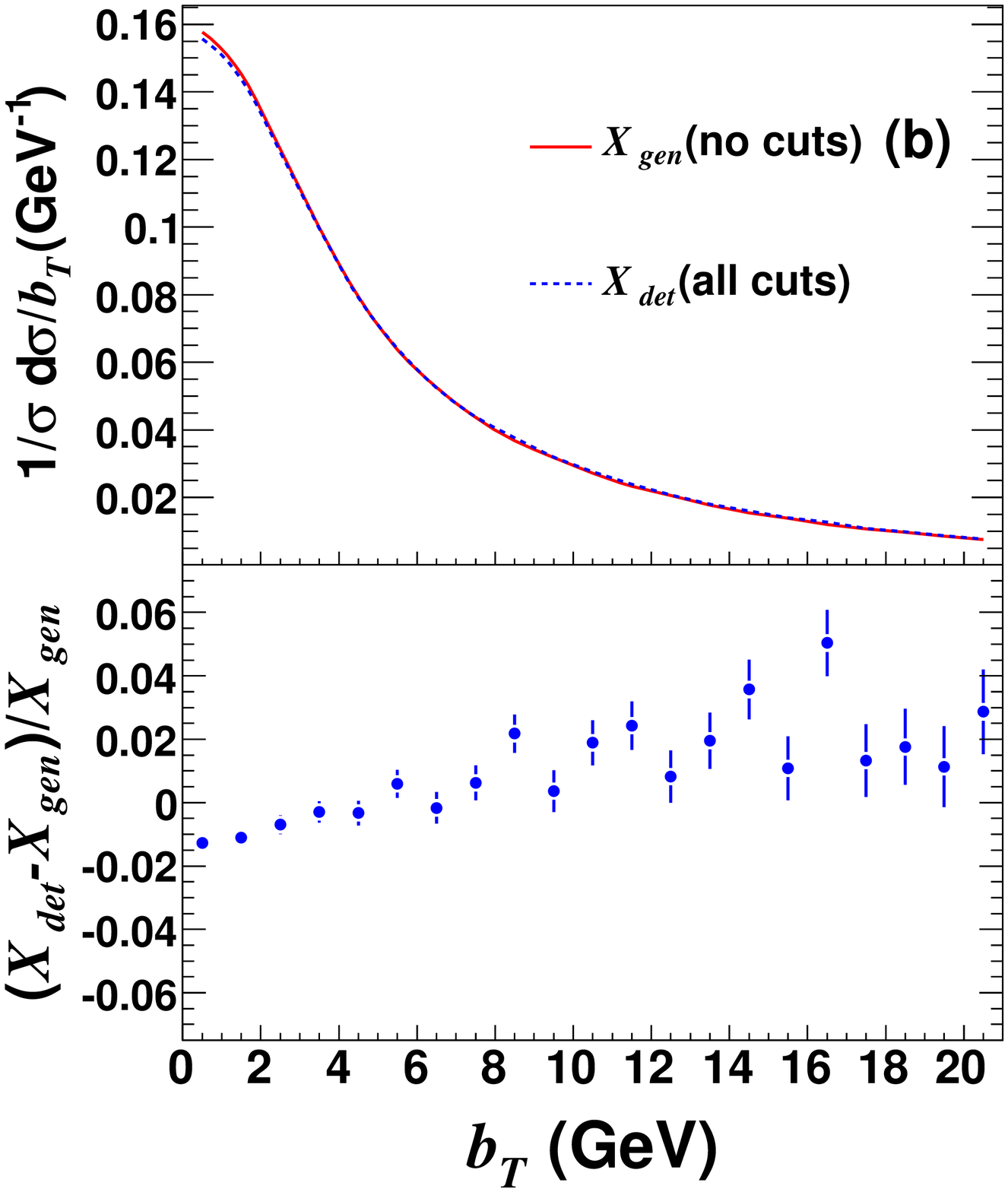}
    \includegraphics[width=0.49\textwidth]{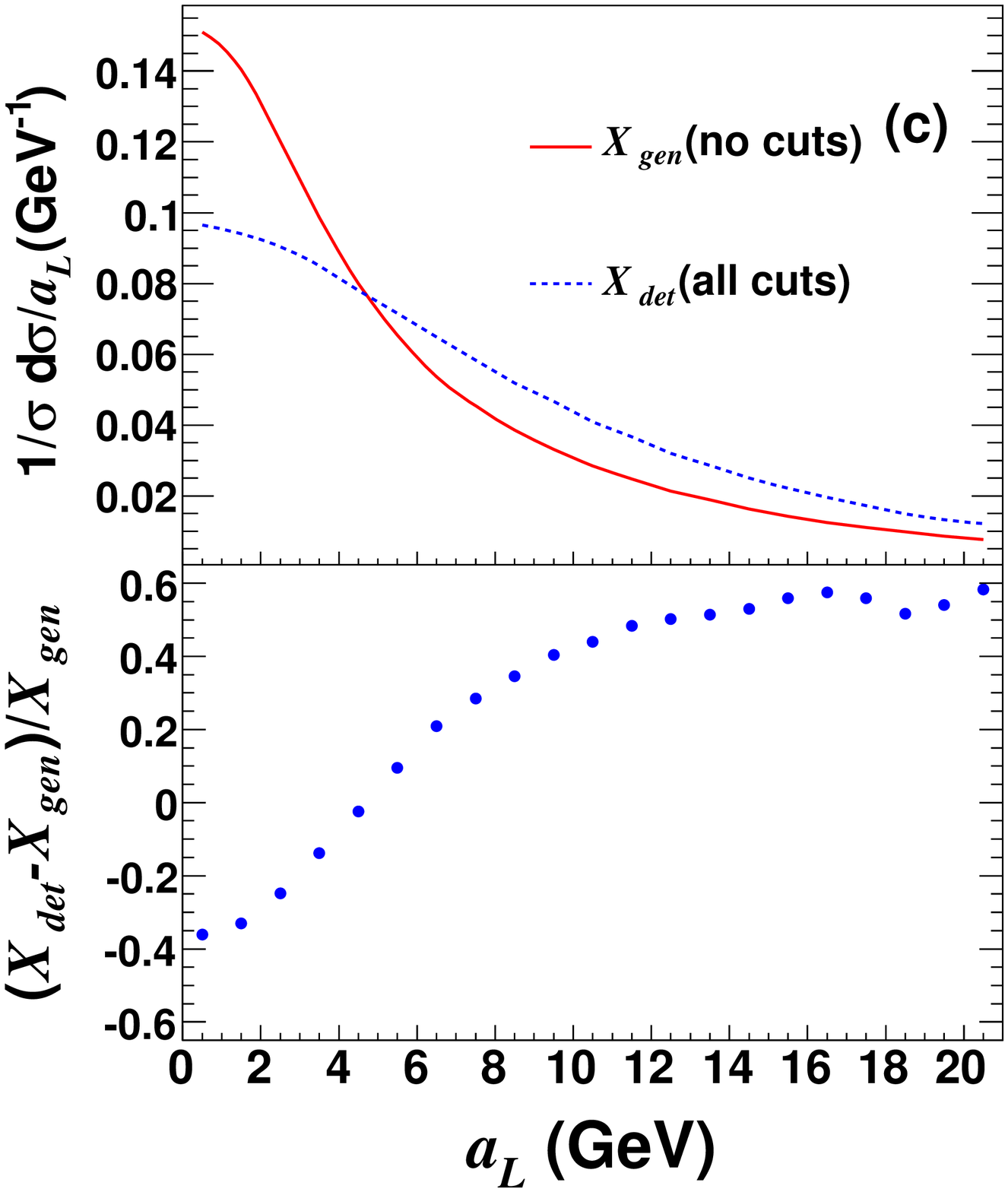}
    \includegraphics[width=0.49\textwidth]{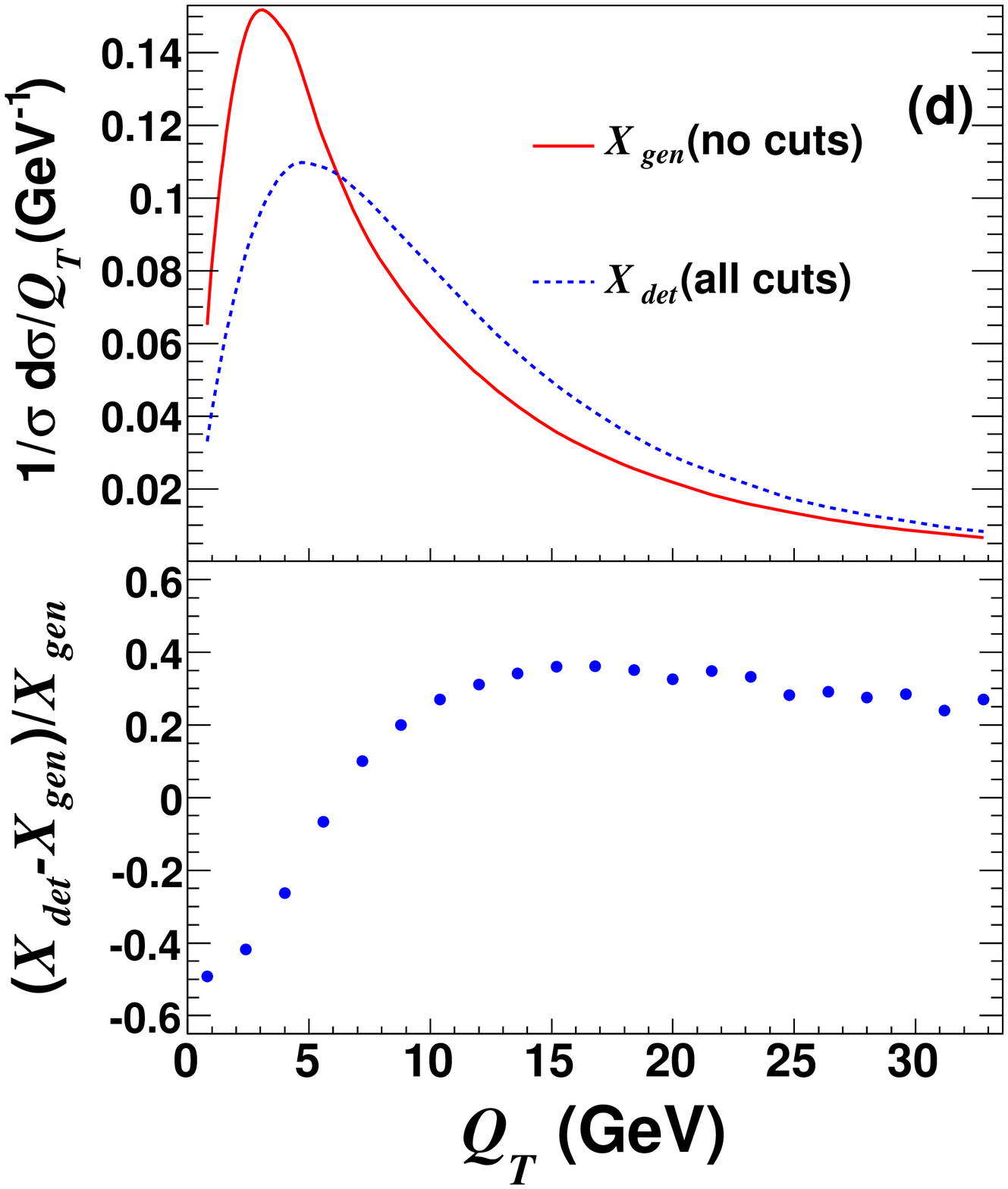}
    \caption[Caption for the figures page]{
      Detector and generator level distributions for (a) \at, (b) \bt, (c)
\al\ and  (d) \zpt. 
      The detector level distributions are for Gaussian smearing in $1/p_T$
of width 0.003 GeV$^{-1}$,
      and all selection cuts are applied. The generator level
distributions do not include selection cuts.
      The lower halves of each plot show the fractional differences.
    }
    \label{smearing}
  \end{figure*}

  Figure~\ref{smearing} compares (separately for \zpt, \at, \al\ and \bt)
 the {\it generator level} distributions {\it without} selection cuts and
  the {\it detector level} distributions {\it with} selection cuts. 
  The detector level \zpt\ and \al\ distributions are substantially affected by the detector resolution and event selection efficiency.
  In contrast the distributions in \at\ and \bt\ are almost completely unaffected.

  In order to extract the underlying shape, the measured distributions would need to be unfolded for the
  detector resolution and corrected for the event selection efficiency with associated systematic uncertainties.
  These uncertainties would be substantially smaller for \at\ or \bt\ than for \zpt\ or \al.
  In fact, for any realistic detector resolution, the measured \at\ or \bt\ distribution describes the
  underlying distribution within a few percent without any unfolding or efficiency correction at all.
  The statistical sensitivity of \at\ or \bt\ to the shape of the underlying \zpt\ distribution is also enhanced by the lack of resolution
  smearing compared to \zpt.
    
  \section{Uncertainties in Measurements of NP Phenomenological Parameters}

\begin{figure}  
\centering 
\includegraphics[width=0.8\textwidth]{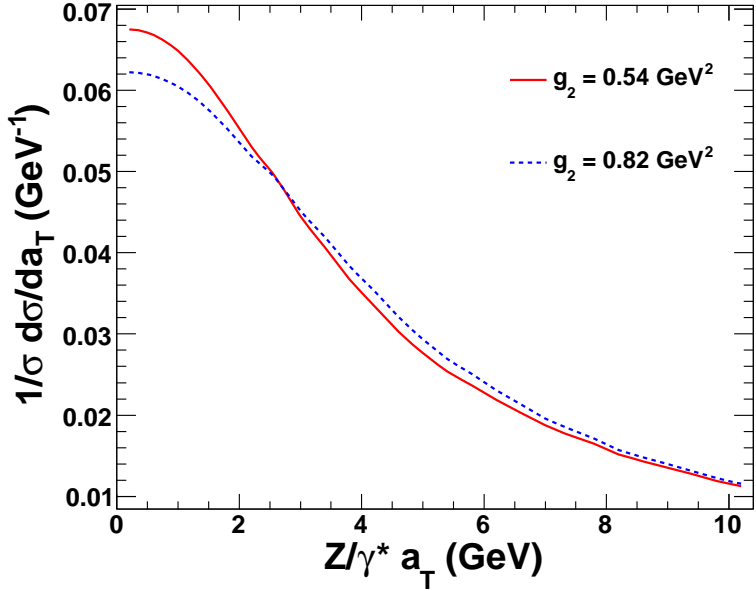}    
\caption{The generator level prediction of the \at\ distribution
of {\sc pythia} (re-weighted to ResBos) for two different $g_2$ values.}                           
\label{at_diff_g2}        
\end{figure} 

  In a ``toy'' MC measurement of the BLNY parameter $g_2$, 
  we study both the sensitivity to systematic uncertainties and the statistical sensitivity
  of \at, \al, \zpt\ and \bt.
   Figure~\ref{at_diff_g2} shows the {\sc pythia} (re-weighted to ResBos) prediction of the normalized \at\   
  distribution for two different values of $g_2$.          
    The \at\ distribution is clearly sensitive to the value of $g_2$.

 Event samples are generated using ResBos, for fifteen $g_2$ values from 0.54 to
0.82 (distributed
 around the world average, $g_2$ = 0.68~$^{+0.02}_{-0.01}$ GeV$^2$)~\cite{BLNY2003}.
 Using the re-weighting procedure described earlier, {\sc pythia} ``MC
samples'' are produced corresponding to each of the $g_2$ values.
An independent sample of 200k {\sc pythia} events is re-weighted to the
central $g_2$ value and represents the experimental ``pseudo-data'' sample.

For each of the 15 MC templates, the pseudo-data vs MC $\chi^2$ is
calculated.
The function: $y = a(x-b)^2 + c$ is fitted to the $\chi^2$ as a function of
$g_2$, and a best fit $g_2$ is determined as $b \pm a^{-1/2}$ where the
uncertainty is statistical ($\Delta\chi^2 = \pm 1$).

  This ``toy'' measurement of $g_2$ serves as an example analysis at an experiment.
  Free parameters in any NP model which affect the \zpt\ could in principle be measured using this strategy.

  \subsection{Systematic Uncertainties}

  We study the systematic uncertainties on a fit to $g_2$.
  The following systematic variations are carried out.
  These are typical of the experimental uncertainties expected at a hadron
collider although the size of the variations are chosen to give reasonable
shifts in the fitted $g_2$:

  \begin{itemize}
  \item In the MC, the smearing constant ($\delta(1/p_T)$) is varied
by a factor $\pm$1\% around a central value of 0.003 GeV$^{-1}$.
  \item As a test of sensitivity to mis-measurement of the event selection
  efficiency dependencies on, $|\eta|$, $p_T$ and isolation, events in MC are given $\pm$10\% more weight for each muon with;
  $ 1 < |\eta| < 2$,
  $15 < p_T < 20$ GeV, or
  $1 < f_{iso} < 2.5$~GeV.
\item In the MC, the $\phi$ smearing constant is varied by $\pm$10\%. 
\item As a test of sensitivity to mis-modeling of final state radiation
  (FSR), events in MC are
  given $\pm$10\% more weight if the difference between the generator level $Z$ mass
  and the generator level di-lepton mass is greater than 1 GeV.
  \end{itemize}

    \begin{table}[ht]
     \caption{The shifts (\%) in the fitted $g_2$ for systematic variations in the MC.
     }
    \centering
     \begin{tabular*}{0.8\linewidth}{@{\extracolsep{\fill}}| l | c  c c c| }
      \hline\hline \T \B
			& \at\ & \al\  & \zpt\ & \bt\ \\[0.5ex]
      \hline \T \B
      $p_T$ smearing $\pm$1\%   & $-0.02$ & $\mp 11$ & $\mp 2.8$ & $\mp 0.03$\\ [0.5ex]
      \hline \T \B
      $p_T$ $\pm$10\%           & $\mp 0.2$ & $\pm 3.8$   & $\pm 1.4$ & $\mp 0.2$\\ [0.5ex]
      \hline \T \B
      $f_{iso}$ $\pm$10\%       & $\mp 0.3$ & $\mp 1.7$  & $\mp 0.59$ & $\mp 0.3$\\ [0.5ex]
      \hline \T \B
      $|\eta|$ $\pm$10\%        & $\pm 0.4$ & $\pm 11$ & $\pm 3.2$ & $\pm 0.4$\\[0.5ex]
      \hline \T \B
      $\phi$ smearing $\pm$10\% & $-0.01$ &  0.00             &  0.00           & $-0.02$ \\[0.5ex]
       \hline \T \B
      FSR $\pm$10\%             & $\mp 0.4$ & $\mp 0.96 $ & $\mp 0.52$ & $\mp 0.4$ \\[0.5ex]
      \hline\hline
      \end{tabular*}
      \label{g2_syst}
     \end{table}

  Table~\ref{g2_syst} shows the shifts in the fitted $g_2$ for the +ve and
  -ve systematic variations.
  For each of the variations, the shift in the fitted $g_2$ is substantially
  smaller using \at\ or \bt\ than \zpt,
  and larger using \al. 
  In fact the only variation to which \at\ or \bt\ are more sensitive is the
  $\phi$ smearing, but for any realistic detector $\phi$ resolution, the
  shift in $g_2$ is negligible.
  
  \subsection{Statistical Sensitivity}

  Simply discarding information from \al\ is not optimal in terms of the statistical sensitivity to the shape of the \zpt\ distribution
  (and thus also the value of $g_2$).
  As well as the basic observables, \zpt, \at\ and \al,
  the following ideas are proposed as possible optimised combinations of $a_T$ and $a_L$:
  \begin{itemize}
  \item A weighted quadrature sum of $a_T$ and $a_L$ with more weight ($w$) given to $a_T$:
    \mbox{ $\zptw(w) = \frac{1}{w^2} \sqrt{(w\cdot a_T)^2 + a_L^2}$ }.
  \item A 2D fit to $\frac{d^2\sigma}{da_Tda_L}$.
  \end{itemize}

  \begin{table}[ht]
     \caption{The binning and $1\sigma$ statistical uncertainties for each
     of the $g_2$ fits, for a range of $\delta(1/p_T)$.
     }
    \centering
    \begin{tabular}{|c | c c c c c c|}
      \hline\hline  \T \B
			&  \zpt\ & $a_T$   & $a_L$  & $\zptw(w=5)$ &$\frac{d^2\sigma}{da_T da_L}$  & \bt\  \\ [0.5ex]                                                      
      \hline  \T \B
      nbins             & 20        & 20      & 20     & 20      & 20($a_T$)$\times$20($a_L$)  & 20 \\
      range (GeV)       & 0-30      & 0-20    & 0-20   & 0-30    & 0-20($a_T$), 0-20($a_L$)    & 0-20 \\
      \hline \T \B
      $\delta(1/p_T)$ (GeV$^{-1}$) & \multicolumn{6}{ c|}{ $1\sigma$ statistical uncertainty (\%)}\\ [0.5ex]

      \hline \T \B 
      0.0000            & 1.4       & 2.2     & 2.4    & 1.9     & 1.4   & 2.2          \\
      0.0003            & 1.4       & 2.2     & 2.5    & 1.9     & 1.4   & 2.2       \\
      0.0010            & 1.8       & 2.2     & 3.6    & 1.9     & 1.6   & 2.2 \\
      0.0030            & 3.1       & 2.3     & 8.9    & 2.2     & 2.1   & 2.2       \\	

      \hline\hline
      \end{tabular}
    \label{StatErr}
  \end{table}

  The binning and $1\sigma$ statistical uncertainties for each of the fits
  is presented in Table~\ref{StatErr}.
  Since the relative statistical sensitivity of each of the observables
  depends on the detector resolution,              
  the fit is performed for different widths of Gaussian smearing. 
  All of the fits use equal width bins.
  For ``perfect'' resolution, a fit to \zpt\ and the 2D fit have comparable statistical uncertainties.
  A fit to $a_T$ or $a_L$ alone is less statistically sensitive, as each contains information on only one component of the \zpt.

  At larger $\delta(1/p_T)$, the sensitivity of $a_L$ is completely washed out by the smearing.
  For $\delta(1/p_T) = 0.003$ GeV$^{-1}$, $a_T$ alone gives better statistical precision than \zpt.
  Over the resolution range covered, $\zptw(w=5)$ gives slightly better statistical precision than $a_T$ alone.
  Note that no attempt has been made to optimize the weight in $\zptw(w=5)$ as a function of resolution, 
  and $w = 5$ is a somewhat arbitrary choice.
  Clearly, for ``perfect'' resolution, $w = 1$ is the only sensible choice.
  Once experimental systematic uncertainties are taken into account,
  $a_T$ will need to be given more weight in any optimal combination of $a_T$ and $a_L$.

  \subsection{Sensitivity to Small-x Broadening}

  The NP form factor (Eq.~\ref{BLNYeq}) required an alteration,
  to describe deep inelastic scattering (DIS) data involving initial state partons with
  $x < 10^{-3}$~\cite{Nadolsky-smallx-2001}.
  Berge {\sl et al.}~\cite{BergeSmallx-2005}~suggested that if this (so-called small-$x$ broadening) was observed at the
  Tevatron in an exclusive high $|y|$ sample of $Z$ bosons,
  a broader Higgs (and $W,Z$) boson transverse momentum distribution could be expected at the LHC.
  The D\O\ Run~II data on \zpt\ \cite{DzeroRunIIa} disfavoured this modification, although given the large uncertainties,
  the data was not particularly sensitive to such broadening.
  Even without taking into account the reduced systematic uncertainties,
  the optimised fits (\zptw, $d^2\sigma/da_T da_L$) described earlier would be more sensitive to such effects.
  Once systematic uncertainties are taken into account, the sensitivity is further enhanced relative to the
  \zpt.  

  \subsection{Azimuthal Correlation Between the Recoil and the
Leptonic Decay}
  The example $g_2$ measurement presented would be sensitive to the description by the MC event
  generator(s) of                                                       
  the azimuthal correlation between the hadronic recoil and the leptonic
  decay.                                                         
  A measurement of $d^2\sigma/da_Tda_L$ could be used to study this
  correlation. 

  \section{Conclusions}
  Using MC simulations we demonstate the potential benefits of decomposing
  the \zpt\ into two orthogonal components, \at\ and \al.
  A measurement of the \at\ distribution would be substantially less sensitive to two of the dominant 
  experimental systematic uncertainties reported in previous \zpt\
  measurements: lepton $p_{T}$ or $E_{T}$ mis-measurement,
  and the \zpt\ dependence of the event selection efficiencies.
  A slightly different variable \bt\ is demonstrated to be similarly
  insensitive to these uncertainties.
  An optimal combination of $a_T$ and $a_L$, giving more weight to $a_T$ gives,
  for any realistic detector resolution, better statistical sensitivity to the shape of the region of low \zpt.
  Two possibilities are proposed: a weighted quadrature sum of $a_T$ and $a_L$ or a fit to $d^2\sigma/da_T da_L$.

  A measurement of $d^2\sigma/da_T da_L$ could potentially probe the azimuthal correlation between
  the $Z$ boson decay axis and the hadronic recoil.
  The partial differential cross sections; $d^2\sigma/da_T dQ$, $d^2\sigma/da_T dy$ and 
  $d^3\sigma/da_T dQ dy$ could also be measured, taking advantage of the reduced systematic 
  uncertainties on $a_T$ compared with similar measurements of differential cross sections with respect to \zpt.
  Such distributions would be sensitive to small-$x$ broadening effects.


 





\end{document}